\documentclass{PoS}

\graphicspath{{./Figures/}}

\leftline{\parbox{11cm}{\large\rm DESY~09-199, HU-EP-09/58, MS-TP-09-26, SFB/CPP-09-112}}
\vspace*{-0.5cm}

\title{On the phase structure of lattice QCD with 
twisted-mass Wilson fermions at non-zero temperature}

\ShortTitle{Phase structure of twisted-mass lattice QCD at $T \ne 0$}

\author{The tmfT Collaboration:
\thanks{Supported by DFG via SFB/TR9 ``Computational Particle Physics''}
}
\author{
  \speaker{Michael M\"uller-Preussker} \\
  Humboldt-Universit\"at zu Berlin, Institut f\"ur Physik, D-12489 Berlin, Germany \\
  E-mail: \email{mmp@physik.hu-berlin.de}
}
\author{  
  Ernst-Michael~Ilgenfritz \\
  Humboldt-Universit\"at zu Berlin, Institut f\"ur Physik, D-12489 Berlin, Germany \\
  Universit\"at Heidelberg, Institut f\"ur Theoretische Physik, 
          D-69120 Heidelberg, Germany \\
  E-mail: \email{ilgenfri@physik.hu-berlin.de}
}
\author{
  Karl Jansen$~^{*}$ \\
  NIC, DESY, D-15738 Zeuthen, Germany \\ 
  E-mail: \email{Karl.Jansen@desy.de}
}
\author{
  Maria Paola~Lombardo \\
  Laboratori Nazionali di Frascati, INFN, I-100044 Frascati, Roma, Italy \\
  E-mail: \email{Mariapaola.Lombardo@lnf.infn.it}
}
\author{
  Owe~Philipsen and Lars~Zeidlewicz \\ 
  Universit\"at M\"unster, Institut f\"ur Theoretische Physik, 
  D-48149 M\"unster, Germany, \\ 
  E-mail: \email{O.Philipsen@uni-muenster.de} \\
  E-mail: \email{Zeidlewicz@uni-muenster.de}
}
\author{
  Malik Kirchner, Marcus~Petschlies, David Schulze, and Carsten Urbach\\
  Humboldt-Universit\"at zu Berlin, Institut f\"ur Physik, D-12489 Berlin, Germany \\
  E-mail: \email{Malik.Kirchner@physik.hu-berlin.de} \\
  E-mail: \email{Marcus.Petschlies@physik.hu-berlin.de} \\
  E-mail: \email{David.Schulze@physik.hu-berlin.de} \\
  E-mail: \email{Carsten.Urbach@physik.hu-berlin.de} 
}

\abstract{In this talk we give an overview of the 3D phase diagram of two-flavour 
non-zero temperature lattice QCD with twisted-mass Wilson fermions and a tree-level 
Symanzik-improved gauge action. We present a first feasibility study at maximal 
twist and, for the quenched case, we demonstrate automatic $\mathcal{O}(a)$-improvement 
to work.
\vspace*{1.0cm}
}

\FullConference{The XXVII International Symposium on Lattice Field Theory\\
		 July 26-31, 2009\\
		 Peking University, Beijing, China}

\begin{document}

\section{Introduction}
The aim of this work was to explore the applicability of the  
twisted-mass Wilson fermion formulation \cite{Frezzotti:2000nk,Frezzotti:2003ni}
as described in the review
by A. Shindler~\cite{Shindler:2007vp} for investigations of lattice QCD at 
non-zero temperature. The use of the staggered-fermion formulation has 
computational advantages~\cite{Jansen:2008vs}, 
but remains conceptually controversial~\cite{Creutz:2007rk}.
On the other hand, the often used clover-improved Wilson fermion formulation 
requires to determine action and operator specific 
improvement coefficients.  
The twisted-mass formulation, combined with a tree-level Symanzik-improved gauge 
action~\cite{Boucaud:2007uk,Boucaud:2008xu}, appears to be a challenging 
alternative for non-zero temperature lattice simulations, since it offers 
automatic $\mathcal{O}(a)$-improvement by tuning the bare quark mass 
parameter only. It allows high-statistics simulations in the range of 
pion masses $m_{\pi} \gtrsim 270 {\rm MeV}$.

As a first step we had to characterize the phase structure of the model by 
locating the transition/crossover lines and surfaces in the three-dimensional 
$\beta-\kappa-\mu_0$-space. The results of this study supporting a conjecture
for the phase diagram by M. Creutz from the chiral perturbation theory point of 
view~\cite{Creutz:1996bg,Creutz:2007fe} were already presented in 
Refs.~\cite{Ilgenfritz:2008td,Ilgenfritz:2009ns}. Here we give an overview of the
phase diagram but concentrating on the thermal transition surface. Moreover,
we discuss a first feasibility study carried out at maximal twist for large
quark mass. In the quenched case we are going to demonstrate that automatic 
$\mathcal{O}(a)$ improvement also works in the finite-temperature case.

\section{The 3D phase diagram and the thermal transition}
\begin{figure}[ht]
\begin{center}
\includegraphics[height=6.0cm,width=5.5cm,angle=-90]{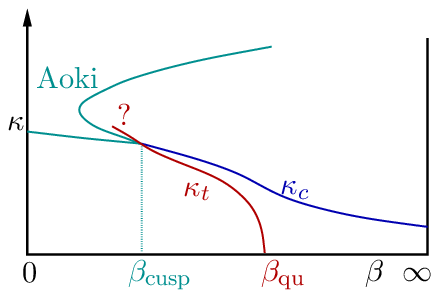} $\qquad$
\includegraphics[height=6.5cm,width=6.5cm,angle=-90]{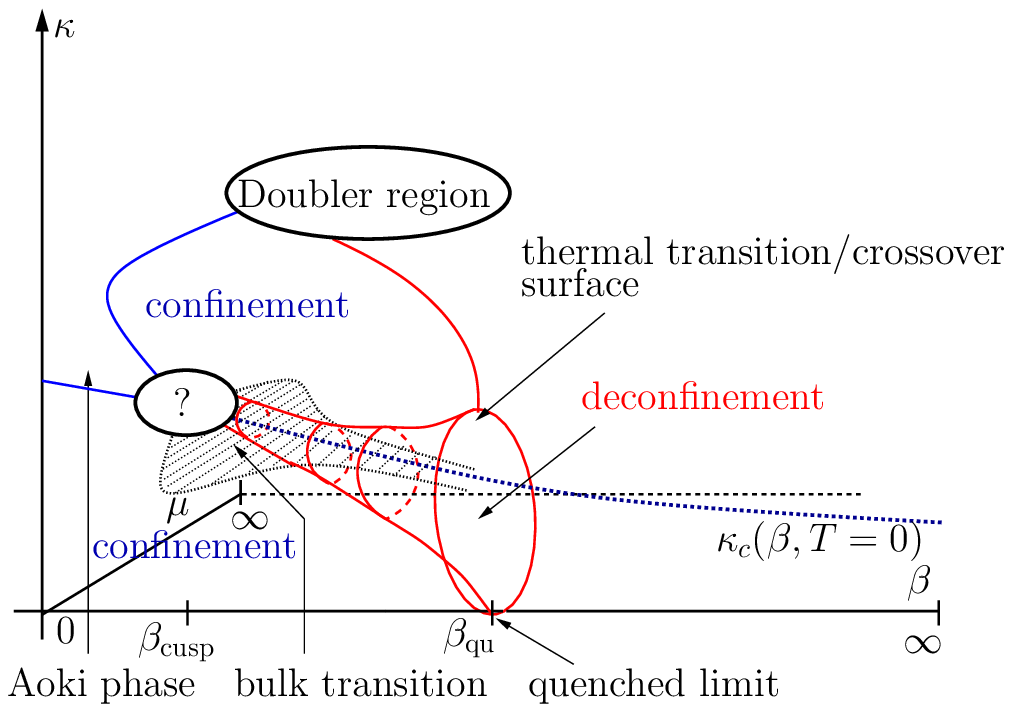}
\end{center}
\vspace{-0.5cm}
\caption{Schematic view of the phase structure as seen in older 
investigations~\cite{AliKhan:2001ek} for a temporal lattice 
extent $N_{\tau}=4,6$ (left) and as found in this work~\cite{Ilgenfritz:2009ns}
with twisted-mass fermions in the $\beta-\kappa-\mu_0-$diagram for 
$N_{\tau}=8$ (right).}
\label{fig:overview}
\end{figure}

The $\beta-\kappa$-phase diagram for two-flavour lattice QCD with clover-improved 
Wilson fermions has been thoroughly studied for small time-extent $N_{\tau}=4,6$ 
a few years ago by the CP-PACS collaboration~\cite{AliKhan:2000iz,AliKhan:2001ek}. 
A schematic view of the emerging phase structure is shown in the left panel of 
Fig. \ref{fig:overview}. The cusp of the strong coupling Aoki phase~(see 
\cite{Ilgenfritz:2003gw,Ilgenfritz:2005ba} and references therein) 
-- the latter (in the infinite-volume limit) characterized by a non-vanishing 
expectation value $ \langle \overline{\psi} i \gamma_5 \tau^3 \psi \rangle$ 
indicating the spontaneous breakdown of a combined parity-flavour symmetry -- 
seemed tightly connected with the thermal transition line $\kappa_t(\beta)$. 

Here, we consider Wilson fermions with the 
additional twisted mass term $~\mu_0~\overline{\psi} i \gamma_5 \tau^3 \psi$. 
For the gauge action the tree-level Symanzik-improved gauge action is employed. 
Having included the twisted mass term a more complicated 3D phase diagram
has to be explored. For lattice sizes $N_{\tau}=8, N_{\sigma}=16$,  
we were able to show~\cite{Ilgenfritz:2009ns} that the Aoki phase 
ends somewhere inside the interval $\beta=3.0,\ldots,3.4$ and, around $\beta=3.4$, 
becomes replaced by a region of metastabilities indicating a first order 
transition area (the shaded area in the right panel of Fig. \ref{fig:overview}), 
a remnant of a transition known also in the zero-temperature 
case~\cite{Sharpe:2004ps,Farchioni:2004us,Munster:2004am,Farchioni:2004fs}. 

\begin{figure}[ht]
\begin{center}
\includegraphics[height=5.5cm,width=4.5cm,angle=-90]{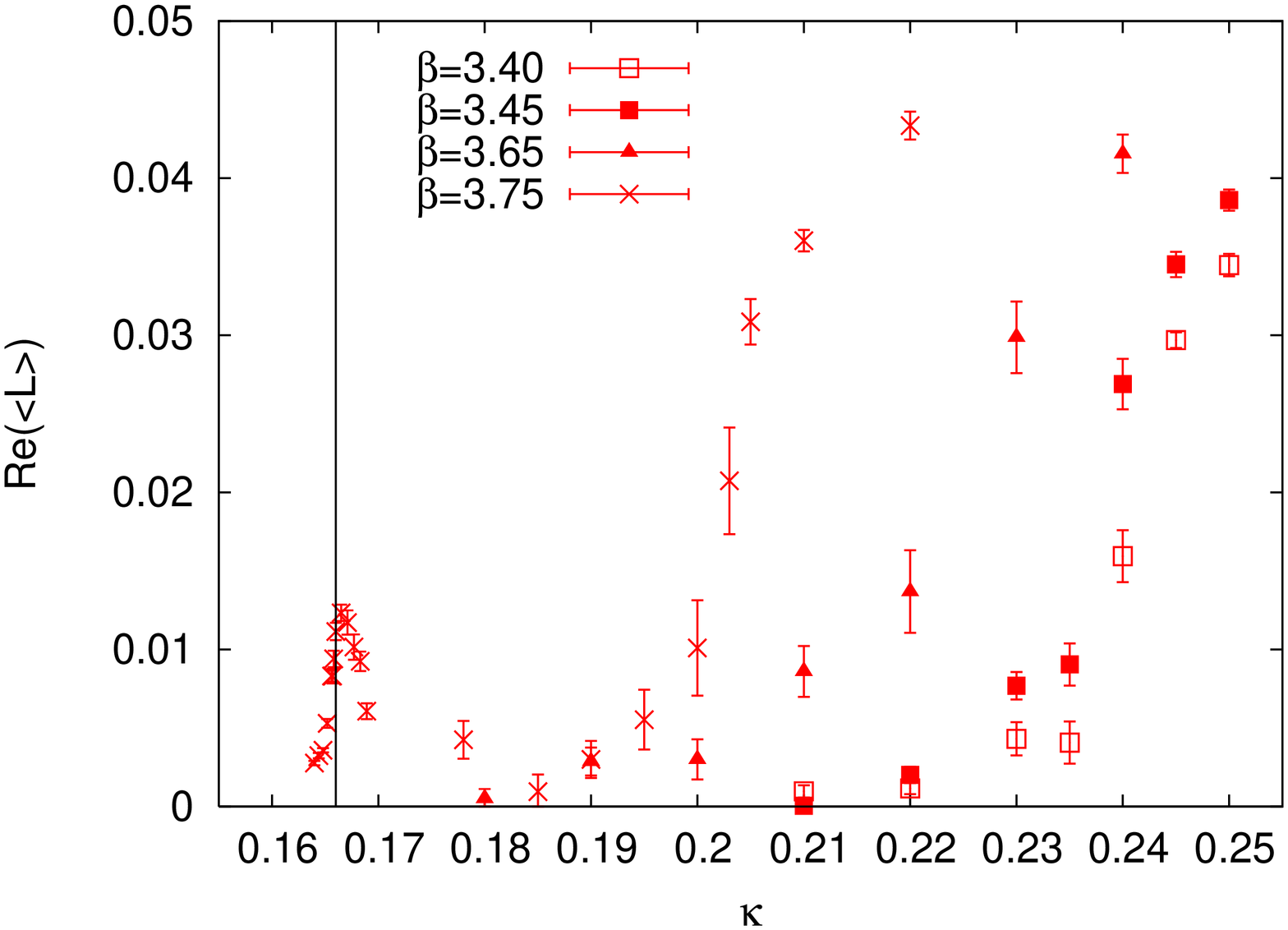} $\qquad$
\includegraphics[height=5.5cm,width=4.5cm,angle=-90]{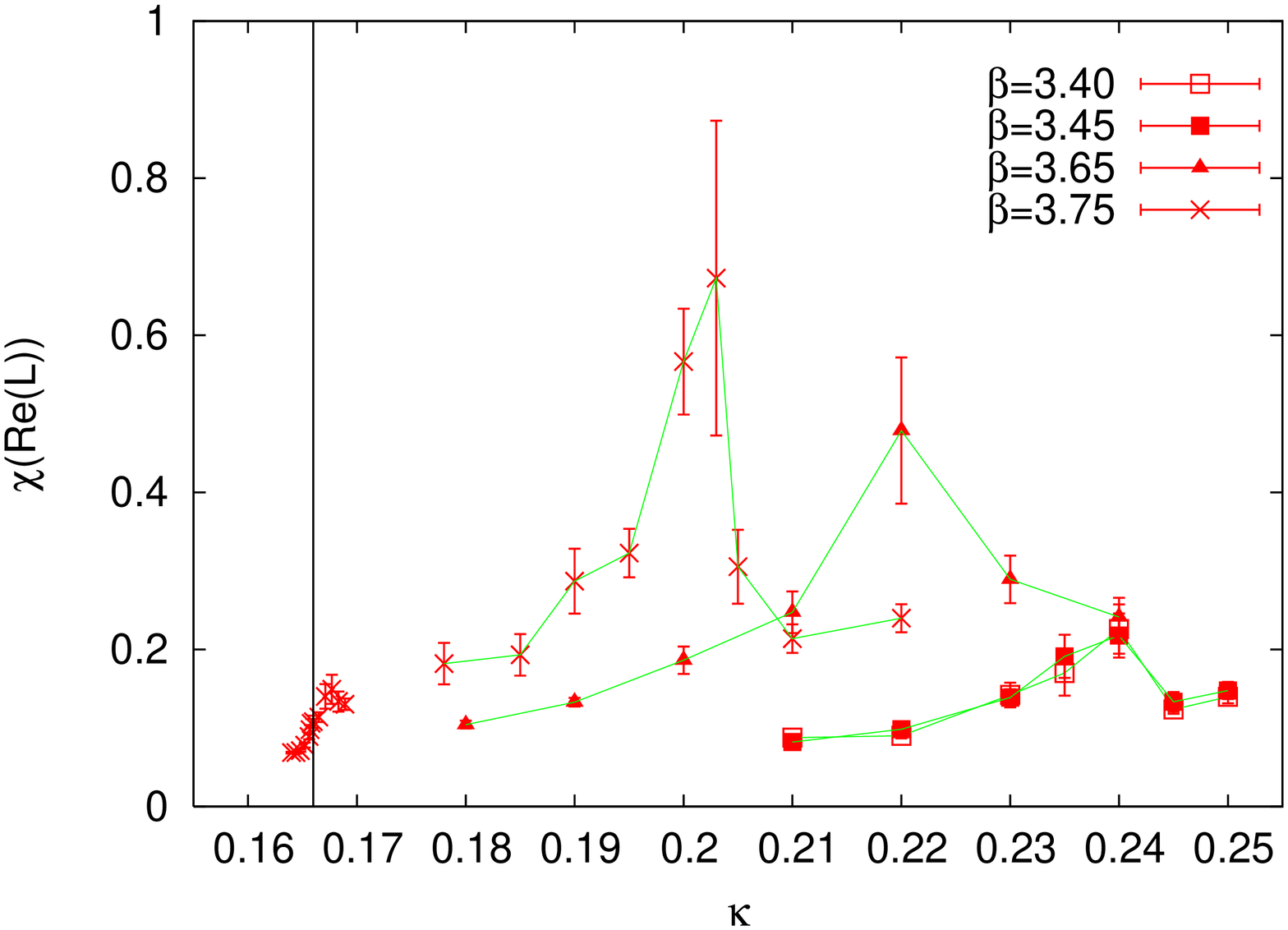}
\end{center}
\vspace{-0.5cm}
\caption{$\kappa$-scans of the Polyakov loop (left) and Polyakov loop 
susceptibility (right) for various $\beta$-values ($\beta=3.4, 3.45, 3.65$ for
$\mu_0=0.0068$; $\beta=3.75$ for $\mu_0=0.005$). Vertical lines mark
$\kappa_c(\beta=3.75)$. }
\label{fig:polyakov}
\end{figure}

In what follows we are concentrating on the thermal transition seen at  
values $\beta \gtrsim 3.65$ and not too small $\mu_0$ (otherwise we are still 
running into the metastability region). Since the hopping parameter 
$\kappa$ and the twisted mass parameter $\mu_0$ are directly connected 
with the bare quark mass
\begin{equation}
 m_q = \sqrt{\frac{1}{4}\left(\frac{1}{\kappa}-\frac{1}{\kappa_c}\right)^2 + 
 \mu_0^2 }\,, 
\label{eq:mass}
\end{equation}
we expect a cone-like structure of surfaces of equal physics around the critical
chiral line $\kappa=\kappa_c(\beta), ~\mu_0=0$. As a first step one can scan
the phase diagram in a larger $\kappa$-range in order to see how the thermal
transition surface extends above $\kappa_c(\beta)$. The result is shown
in Figs. \ref{fig:polyakov}. For $\beta$-values $\beta=3.4, 3.45, 3.65, 3.75$ 
from the steep rises of the Polyakov loop and from maxima of its susceptibility 
we observe very clear signals for a thermal transition in $\kappa$. 
But additionally, for $(\beta=3.75, \mu_0=0.005)$, we see a tiny 
$\kappa$-interval around $\kappa_c=0.166$ where the Polyakov loop exhibits a 
comparably little maximum, which could have been easily overlooked. Thus,  
with rising $\kappa$ starting from values below $\kappa_c$ 
we pass through subsequent confinement-deconfinement, deconfinement-confinement 
transitions (or better crossovers) below and above $\kappa_c$, respectively, 
followed again by a confinement-deconfinement transition far above
$\kappa_c$. The latter transition surface extends to the next fermion doubler 
region in the phase diagram. We have seen by additional $\beta$-scans at fixed
$\kappa > \kappa_c(\beta)$ that the Creutz cone structure~\cite{Creutz:2007fe} 
that we are exploring is connected with the upper confinement-deconfinement 
transition by a phase transition surface bending upward in 
$\kappa$ at larger $\beta$.

\begin{figure}[ht]
\begin{center}
\includegraphics[height=5.5cm,width=4.5cm,angle=-90]{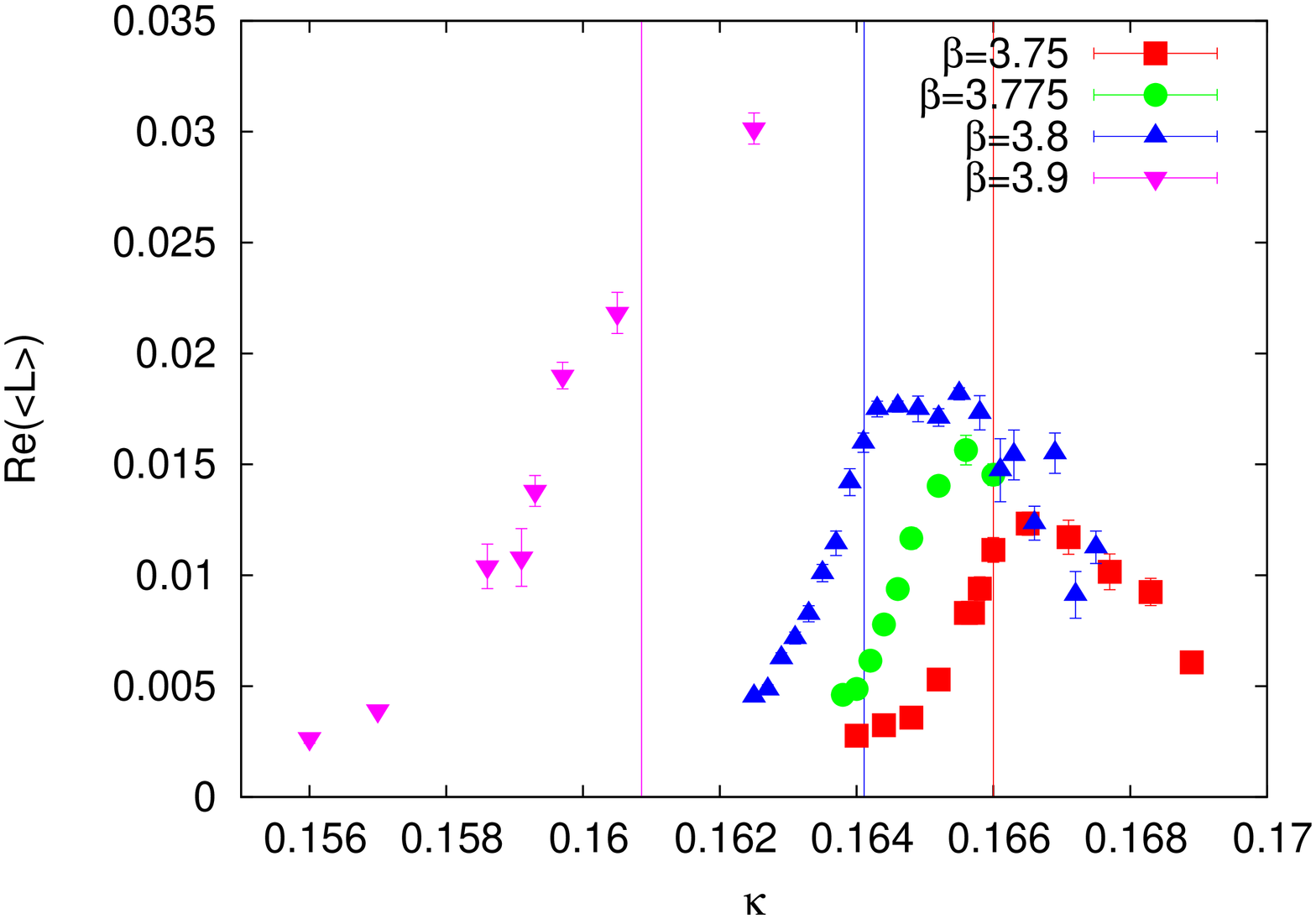} $\qquad$
\includegraphics[height=5.5cm,width=4.5cm,angle=-90]{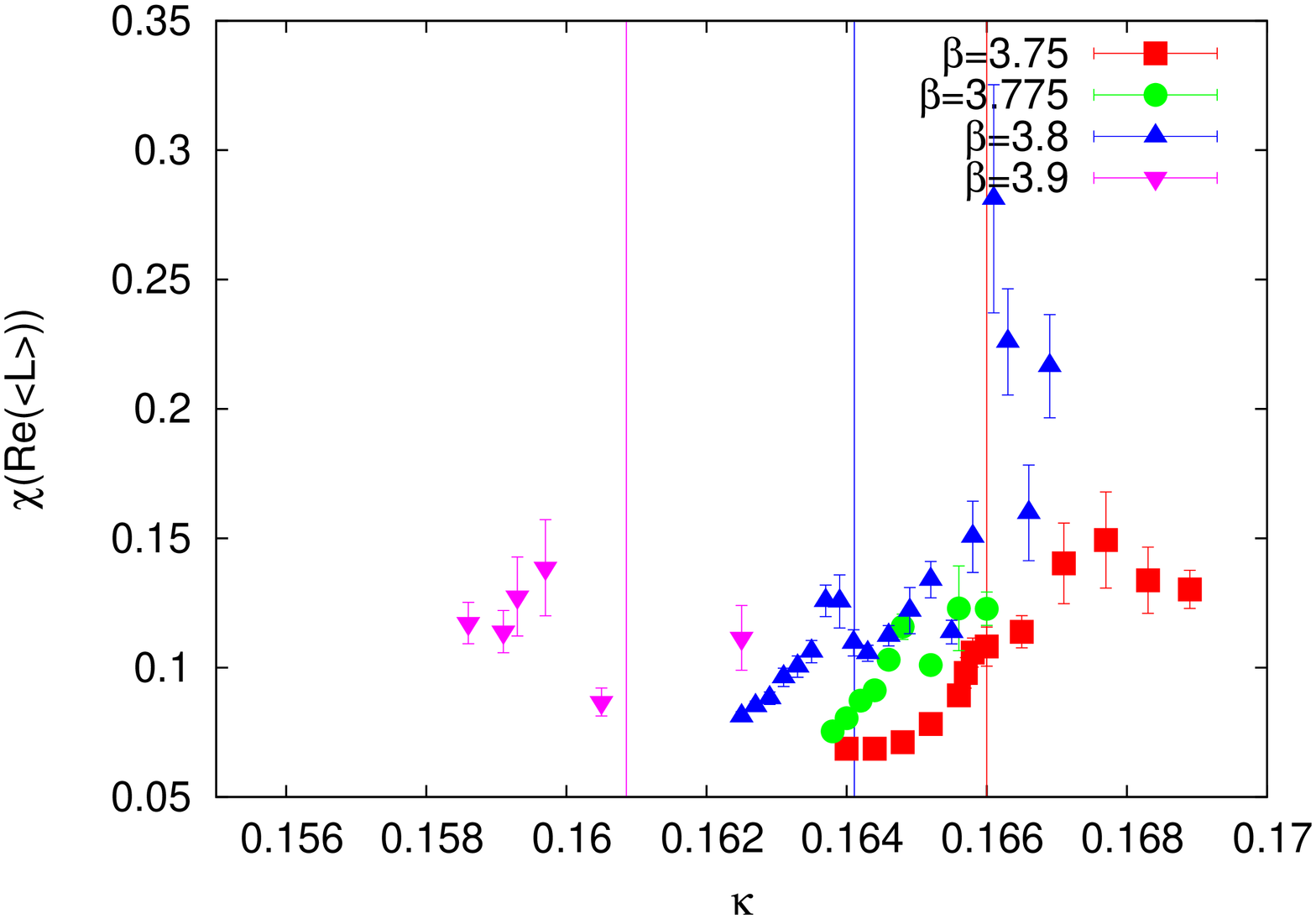}
\end{center}
\vspace{-0.5cm}
\caption{Zoom into the behaviour of the Polyakov loop (left) and its
susceptibility (right) versus $\kappa$ for various $\beta$ and 
$\mu_0=0.005$. Vertical lines from left to right mark the chiral critical
values $\kappa_c(\beta)$ for $\beta=3.9, 3.8$ and $3.75$, respectively.}
\label{fig:polyakov_zoom}
\end{figure}

\begin{figure}[ht]
\begin{center}
\includegraphics[height=4.0cm,width=5.5cm,angle=0]{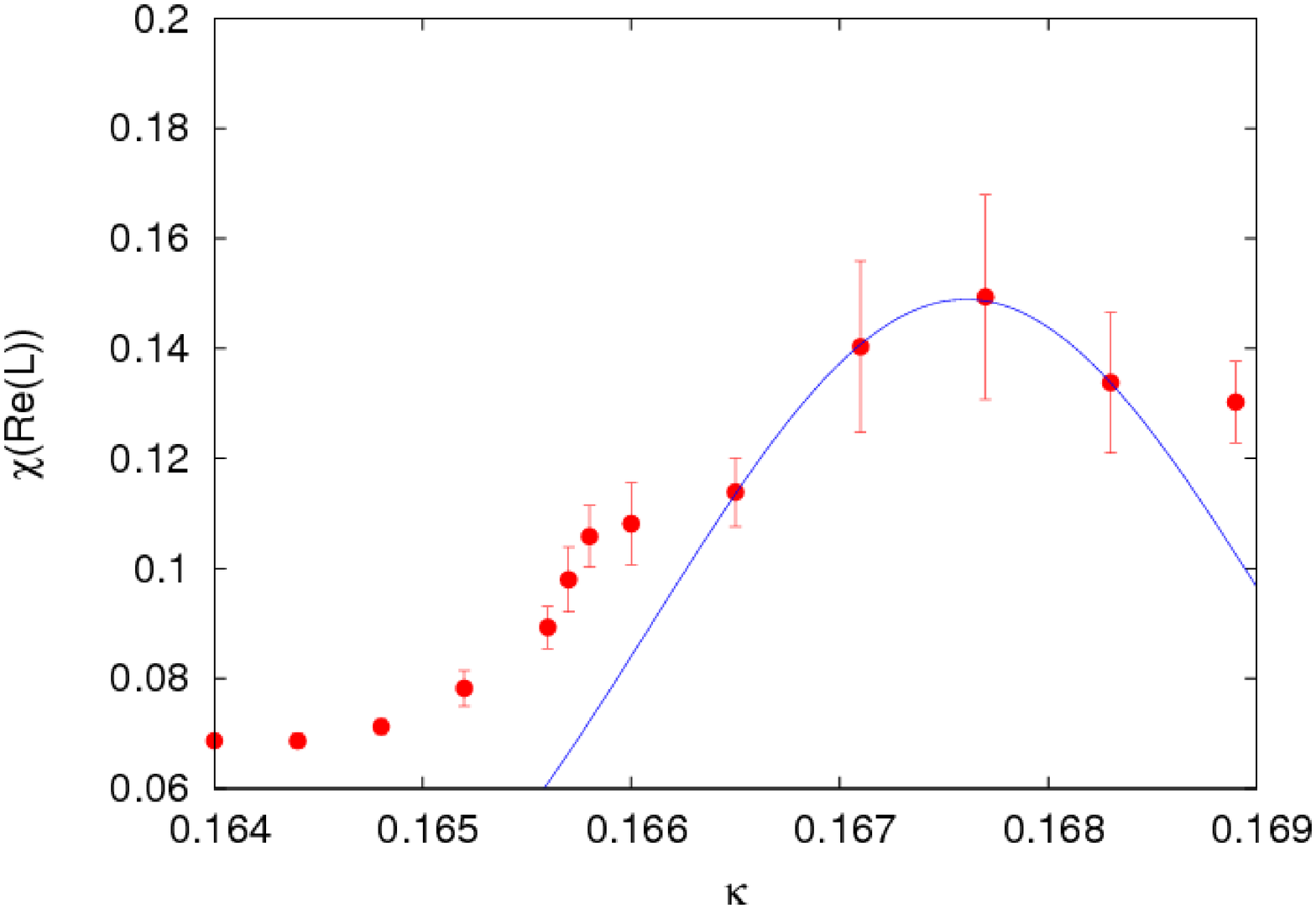} $\qquad$
\includegraphics[height=4.0cm,width=5.5cm,angle=0]{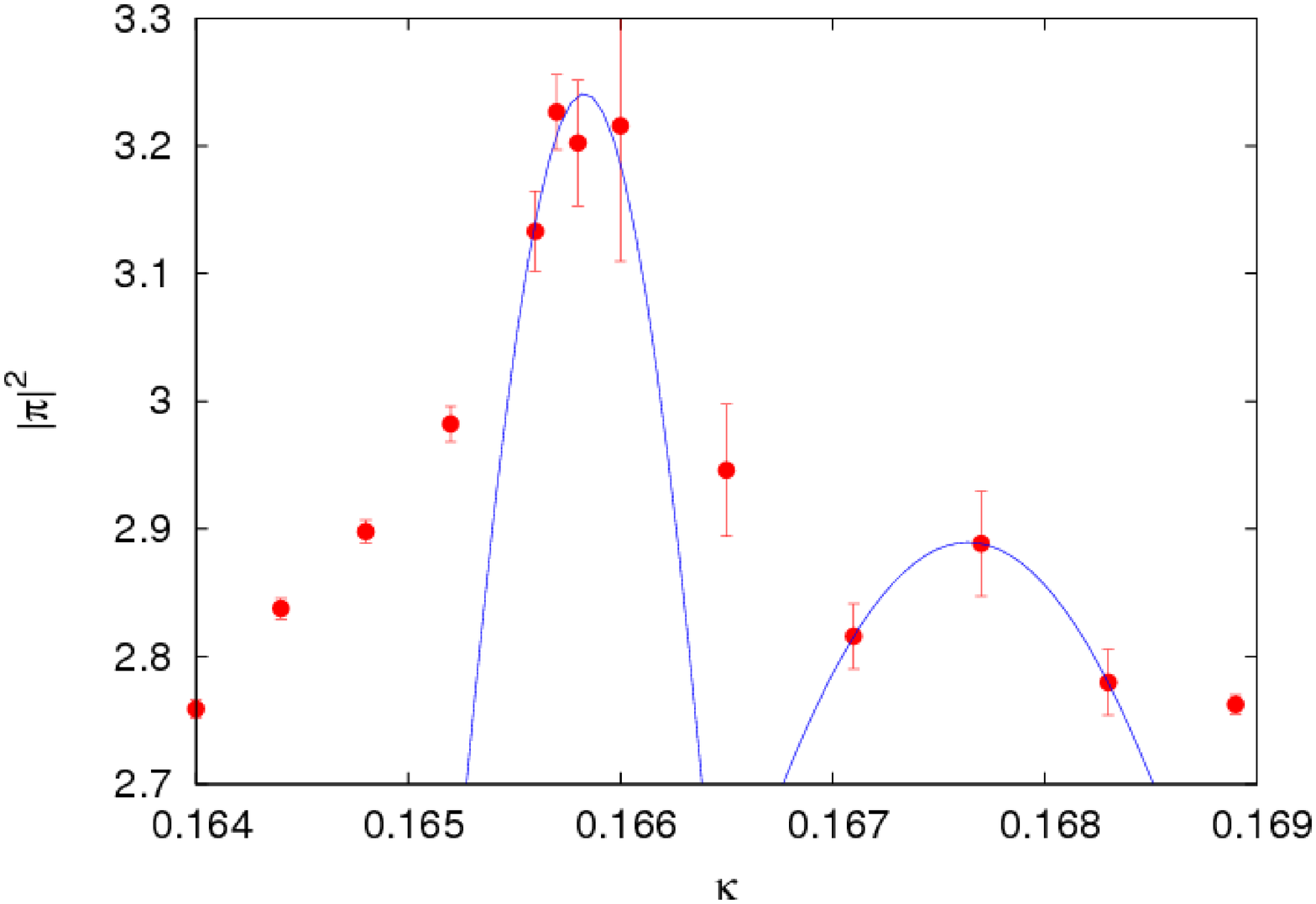}
\end{center}
\vspace{-0.5cm}
\caption{Polyakov loop susceptibility (left) and pion norm (right) versus
$\kappa$ both for $\beta=3.75$ and $\mu_0=0.005$.}
\label{fig:polyakov_pionnorm}
\end{figure}

Zooming into the region around $\kappa_c(\beta)$ the Polyakov loop and its
susceptibility behave as shown in Figs. \ref{fig:polyakov_zoom}. 
The maxima or shoulders of the Polyakov loop susceptibility shown in the 
right panel indicate smooth transitions or crossovers. For $\beta=3.75,
\mu_0=0.005$ this can be clearly seen in Fig. \ref{fig:polyakov_pionnorm},
where also the so-called pion norm has been considered. The Gaussian shape
lines are fitted to highlight the position of the expected crossover.
Note that at the given $\beta$ and $\kappa_c(\beta)=0.166$ the value 
$\mu_0=0.005$ can be related to a pion mass value 
$m_{\pi} \simeq 400 {\rm MeV}$ and a temperature $T \simeq 210 {\rm MeV}$. 
This is close to values recently reported by the DIK 
collaboration~\cite{Bornyakov:2009qh}.

\begin{figure}[ht]
\begin{center}
\includegraphics[height=5.5cm,width=4.5cm,angle=-90]{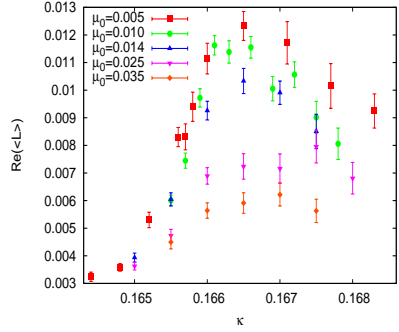} 
\end{center}
\vspace{-0.5cm}
\caption{Polyakov loop scans as function of $\kappa$ at various $\mu_0$ 
for $\beta=3.75$.}
\label{fig:cone_extension}
\end{figure}

We tried to figure out how far the crossover
or transition cone extends in the $\mu_0$-direction. Although we collected 
$O(10^4)$ HMC trajectories per point this turned out to be a quite difficult
task, because of the weak and noisy signals seen in the plaquette and 
Polyakov loop susceptibilities, in the corresponding autocorrelation times 
as well as in the pion norm variable. From $\kappa$-scans for the 
Polyakov loop at $\beta=3.75$ and various $\mu_0$-values drawn in
Fig. \ref{fig:cone_extension} we would like to conclude
that the cone surface ends somewhere in the interval 
$0.014 < \mu_0 < 0.025$. Corresponding fits of the ellipse shape 
distorted by lattice artifacts can be done with an 
expression for the quark or pion mass obtained at next-to-leading order 
in lattice chiral perturbation theory, but still have a quite large
uncertainty~\cite{Ilgenfritz:2009ns}. Therefore, we will not show them 
here.

\section{A first feasibility study at maximal twist}
\begin{figure}[ht]
\begin{center}
\includegraphics[height=5.0cm,width=6.0cm,angle=0]{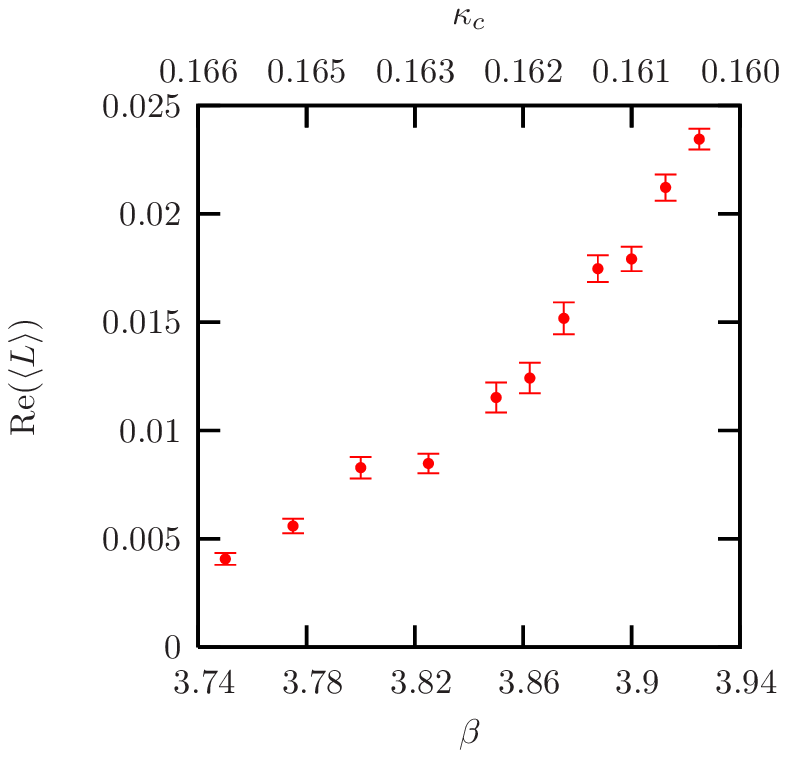} $\qquad$
\includegraphics[height=5.0cm,width=6.0cm,angle=0]{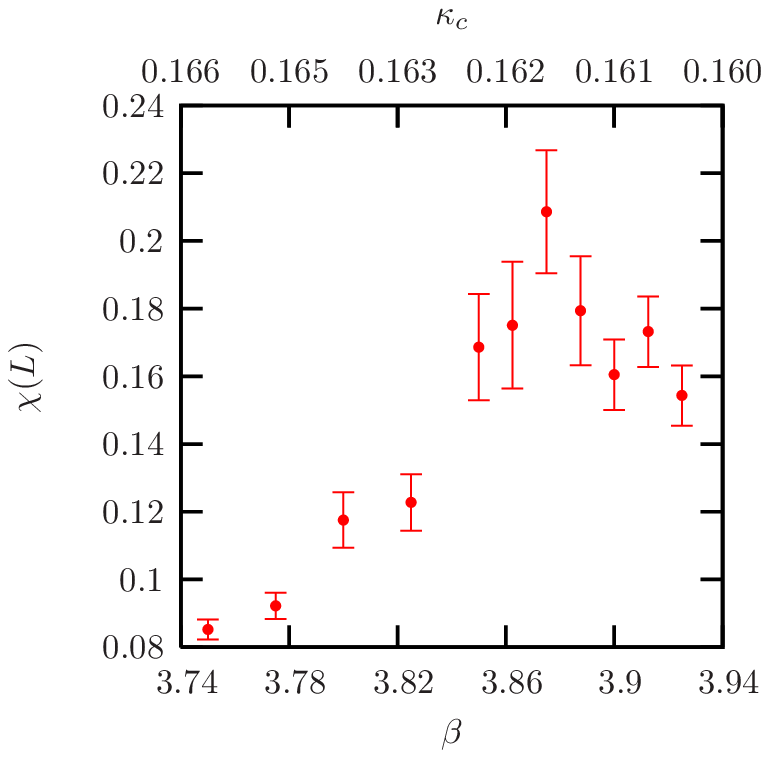}
\end{center}
\vspace{-0.5cm}
\caption{Polyakov loop (left) and its susceptibility (right) versus 
$\beta$ for maximal twist $\kappa=\kappa_c(\beta)$ at $\mu_0=0.040$.}
\label{fig:feasibility}
\end{figure}

So far we have not yet taken advantage of the expected $\mathcal{O}(a)$
improvement. For fixed $\beta$ and $\kappa=\kappa_c(\beta)$ one
would like to change $\mu_0$ in order to vary the physical quark or pion mass.
Since the statistical signals for the crossover turned out to be 
very noisy in this
case, we instead decided to fix $\mu_0$ and to vary $\beta$ and 
$\kappa=\kappa_c(\beta)$ accordingly. The values for $\kappa_c(\beta)$ can be
estimated from the zero-temperature case 
(see e.g. \cite{Boucaud:2007uk,Boucaud:2008xu}\footnote{We acknowledge 
the help of the ETM collaboration providing us also with data prior
to publication.}). For $\mu_0=0.040$ - which corresponds 
to a quite large pion mass value $m_{\pi} \simeq 1 {\rm GeV}$ - we have found the
results shown in Figs. \ref{fig:feasibility}. The `critical' value 
$\beta_t=3.88$ can be translated into $T_c \simeq 280$MeV, which is again
in the same ballpark in comparison with~\cite{Bornyakov:2009qh}.
We conclude that this strategy to satisfy the requirements of an automatic
$\mathcal{O}(a)$ improvement seems to work. 

\section{Automatic $\mathcal{O}(a)$ improvement at $T \ne 0$}
\begin{figure}[ht]
\begin{center}
\includegraphics[height=5.0cm,width=6.0cm,angle=0]{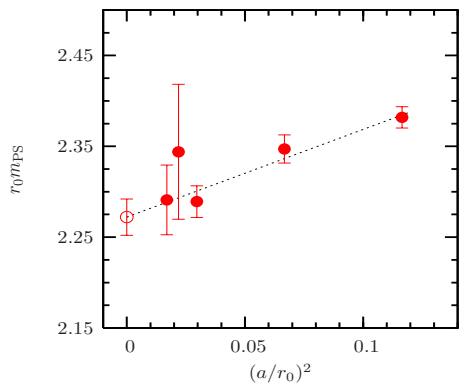}
\end{center}
\vspace{-0.5cm}
\caption{Pseudoscalar screening mass $m_{PS} r_0$ versus lattice spacing
$(a/r_0)^2$ obtained for lattice sizes $N_{\sigma}=24,\ldots,32$ and 
$N_{\tau}=6,\ldots,16$ at fixed $T/T_c = 0.655(5)$ and 
$m_{PS} / m_V \simeq 0.75$.} 
\label{fig:oa_improvement}
\end{figure}

Finally, we have checked that the automatic $\mathcal{O}(a)$ improvement really 
holds in the finite-temperature case. In the quenched approximation we have
computed the pseudoscalar screening mass for varying spatial and temporal 
linear lattice sizes while keeping the physical mass ratio of pseudoscalar 
and vector states and the physical temperature fixed. 
The results are plotted in Fig. \ref{fig:oa_improvement}.
They demonstrate nicely a linear behaviour in the square of the lattice spacing
$~a(\beta)$.  

\section{Conclusions}
We are convinced that with the present study the necessary prerequisites
for a serious non-zero temperature analysis with twisted-mass Wilson 
fermions have been collected in a sufficient manner. The structure 
of the three-dimensional phase diagram has been explored in the physical 
range for the two-flavour case. Still it is difficult to locate the 
(pseudo-) critical behaviour or crossover at fixed 
$(\beta, \kappa=\kappa_c(\beta))$ along the direction 
of varying twisted-mass parameter $\mu_0$. Therefore, in a feasibility study,
we have taken advantage of automatic $\mathcal{O}(a)$ improvement at fixed 
$\mu_0$ by passing through the crossover phenomenon changing $\beta$ 
and keeping close to the chiral critical line $(\beta,\kappa_c(\beta))$ 
for which we can rely on twisted-mass results at zero temperature. 
For the quenched case we have demonstrated that $\mathcal{O}(a)$ 
improvement really works in the non-zero temperature setting . 
We are now in the position to start the determination of the critical 
temperature and of the equation of state with extrapolations 
to the limits of realistic light quark masses and to the continuum. 
In order to reach smaller pion masses we continue our investigation
with $N_{\tau}=10, 12$ on correspondingly larger spatial lattices.

\end{document}